\documentclass[pre,twocolumn,english,superscriptaddress,floatfix,longbibliography]{revtex4-2}
\usepackage{amsmath,amssymb,bm}
\usepackage[a4paper,margin=1in]{geometry}
\usepackage{physics, mathtools, braket}
\usepackage{ulem}
\usepackage{graphicx}
\usepackage{dcolumn}
\usepackage{xcolor}
\usepackage[dvipsnames]{xcolor}
\usepackage{lineno}
\usepackage{hyperref}
\hypersetup{colorlinks=true, citecolor=blue, linkcolor=black, urlcolor=blue}

\begin{document}

\title{Temporal Diffraction Grating for Engineered Superconducting Qubit Dissipation}
\author{Pratik J. Barge}
    \affiliation{Department of Physics, Washington University, St. Louis, Missouri 63130, USA}
    \email{bargep@wustl.edu}
\author{Kater W. Murch}%
    \email{katermurch@berkeley.edu}
    \affiliation{Department of Electrical Engineering and Computer Science, University of California Berkeley, Berkeley, CA, USA, 94720.}
    \affiliation{Department of Physics, University of California Berkeley, Berkeley, CA, USA, 94720}

\date{\today}

\begin{abstract}
Parametric frequency modulation is a standard tool in superconducting circuits for activating tunable interactions and implementing quantum gates. Here, we engineer dissipation in a flux-tunable transmon qubit by using sideband modulation to bring it into resonance with a lossy resonator, opening an on-demand Purcell decay channel. We find that pulsing this channel on and off does not simply lower the time-averaged decay rate; instead, it reorganizes the dissipation spectrum into a structured interference pattern. A Chebyshev-propagator model for the repeated on/off block reproduces the measured spectra and reveals a close structural correspondence to N-slit Fraunhofer diffraction, with each on-window acting as a temporal aperture. By varying the pulse duration and duty cycle, we demonstrate control over the spacing, contrast, and envelope of the dissipation spectrum. These results establish pulsed parametric modulation as a direct method for shaping engineered dissipation in superconducting circuits and provide a new control knob for open quantum system dynamics.
\end{abstract}

\maketitle

\section{Introduction}
Dissipation in quantum systems is often treated as an unwanted source of decoherence, but when structured and controlled it becomes a resource for quantum-state preparation, stabilization, reset, measurement, and simulation \cite{poya96,Verstraete2009,legh15,harrington2022engineered}. In superconducting circuits, one of the most direct forms of dissipation engineering is Purcell engineering: the electromagnetic density of states seen by a qubit can be shaped to suppress spontaneous emission in the protected operating regime, or enhanced to provide a rapid decay channel when reset or cooling is desired \cite{Purcell1946,Reed_2010_PurcellReset,Sunada_2022_IntrinsicPurcellFilter,Swiadek_2024_FastReadoutPurcell,Ding_2025_FastResetProtectiveReadout}. This spectral view of dissipation has been extended beyond simple resonator-mediated decay to deliberately structured reservoirs, including photonic crystals, phononic baths, multimode Purcell filters, and engineered metamaterial environments that tailor the frequency dependence of the loss channel \cite{Harrington_2019_PhotonicCrystalBath,caru20,Kitzman_2023_PhononicBath,Kim_2025_EngineeredBathReset, gu2026multimode}. In this sense, dissipation engineering begins not by eliminating loss, but by determining which transitions are allowed to couple to the environment and which remain protected.

A complementary strategy is to use coherent drives to connect selected states or transitions to lossy auxiliary modes, implementing the logic of optical pumping and sideband cooling in circuit-QED systems \cite{diedrich,vala06,murc12,geer13}. In these protocols, the bath acts as an entropy sink, while drive frequencies and selection rules determine the dark states or manifolds that are left behind. This approach has enabled fast qubit and resonator reset \cite{geer13,bout17,magn18}, autonomous stabilization of entangled states \cite{shan13,lin,kimc16,liu16,ande19,Brown_2022_TradeoffFreeEntanglement,Chen_2025_HardwareEfficientEntanglement}, stabilization of bosonic manifolds and protected states \cite{legh15,lu17,bret15}, and dissipative preparation or stabilization of many-body states \cite{haco15,ma19,PhysRevA.104.032418,Mi_2024_DissipativeManyBody,Pocklington_2022_LocalDissipationVolumeLaw}. Recent work has further emphasized programmable and autonomous forms of reservoir engineering, including autonomous error correction, programmable stabilized states, and energy-selective local reservoirs \cite{Li_2024_AutonomousQEC,Li_2024_LinearCouplerAQEC,Li_2024_ProgrammableStabilization,guo2026entangling}. These examples illustrate a hierarchy of control: the reservoir can be shaped spectrally, accessed selectively by drives, and used to stabilize particular states or manifolds.

Parametric frequency modulation provides a particularly flexible way to access this hierarchy in superconducting circuits. By periodically modulating a qubit, coupler, or auxiliary mode, one can activate otherwise off-resonant interactions through sidebands or parametric resonances, enabling qubit--resonator exchange \cite{Beaudoin_2012_FirstOrderSidebands,Strand_2013_FirstOrderSidebands}, two-qubit gates \cite{McKay_2016_TunableBusGate,Caldwell_2018_ParametricGates,Reagor_2018_UniversalParametricGates,Hong_2019_ACSweetSpotParametricGate,Ganzhorn_2020_ParametricGateNoise,Chu_2020_SuperadiabaticParametricGate,sete2021parametric}, state transfer \cite{Li_2021_ParametricSTIRAP}, fast reset and leakage removal \cite{Zhou_2021_ParametricReset,chen2024fast,Kim_2025_EngineeredBathReset}, and driven dissipation for cavity cooling and reset \cite{Maurya_2024_OnDemandDissipation,Huber_2025_ParametricMultielement}. Most of these applications use continuous or smoothly shaped modulation to activate a desired coherent or dissipative interaction. Here we explore a distinct regime in which a parametrically induced Purcell channel is rapidly switched \textsf{on} and \textsf{off} in time. The resulting dynamics are not described by a simple duty-cycle-averaged decay rate. Instead, repeated access to the lossy resonator produces interference in the dissipative response, reorganizing the Purcell-enhanced decay spectrum into timing-dependent features. Our experiment therefore extends reservoir engineering from spectral selectivity and parametric activation to time-domain shaping of an engineered loss channel, realizing a pulsed Purcell interaction whose dissipative spectrum is controlled by the temporal structure of the modulation sequence.

\begin{figure*}[t]
  \centering
  \includegraphics[width=\linewidth]{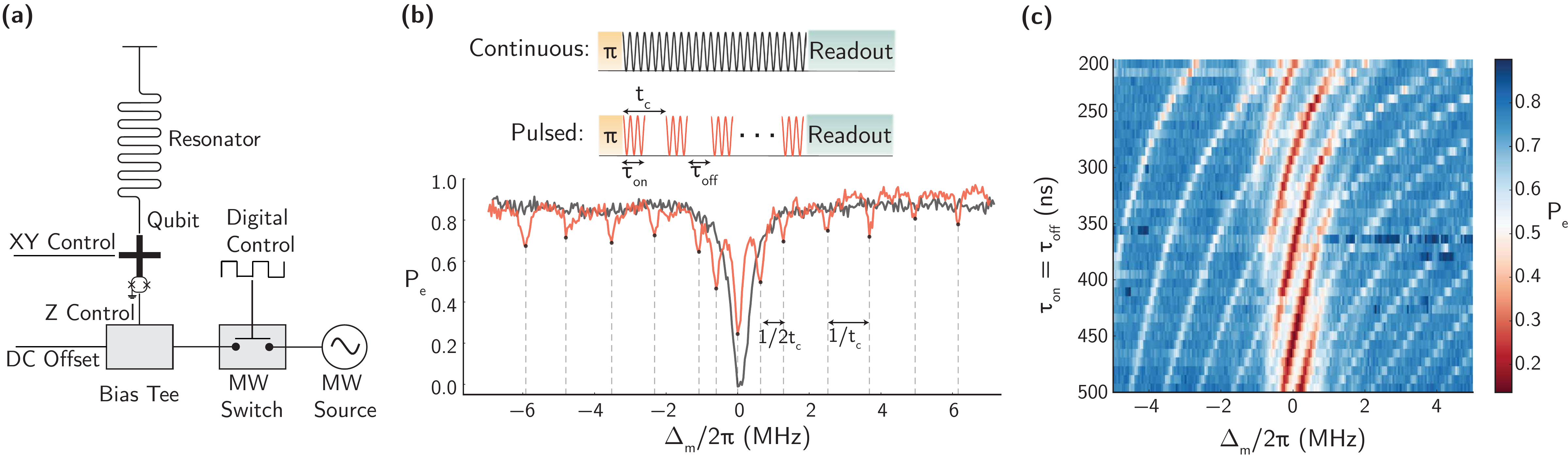}
  \caption{{\bf Experimental protocol and dissipation spectra for continuous and pulsed parametric modulation schemes.} 
  (a) Schematic of the flux-modulation setup. A microwave source is gated by a microwave switch and combined with a DC flux bias through a bias tee before being applied to the qubit via the fast flux line (FFL). (b) Pulse sequences and excited-state probability ($P_e$) for continuous (black) modulation produces a single broad Purcell-enhanced decay feature near resonance, while pulsed (red) modulation produces multiple interference features as a function of modulation detuning $\Delta_{\mathrm{m}}/2\pi$. The pulse sequence consists of repeated \textsf{on} modulation windows of duration $\tau_{\textsf{on}} = 400$ ns separated by equal-duration \textsf{off} windows $\tau_{\textsf{off}}= 400$ ns, followed by qubit readout. (c) 2D plot of $P_e$ for $\Delta_{\mathrm{m}}/2\pi$ and symmetric pulse duration $\tau=\tau_{\textsf{on}}=\tau_{\textsf{off}}$, showing the evolution of the interference pattern with pulse timing.}
  \label{fig:intro}
\end{figure*}

\section{Methods}
\label{sec:methods}
\subsection{Parametric Frequency Modulation and Induced Purcell Decay}

The intrinsic coherence of a transmon qubit is typically protected by operating in the dispersive regime, where the qubit-resonator detuning $\Delta = \omega_{\mathrm{r}} - \omega_{\mathrm{q}}$ is strictly much larger than the static vacuum Rabi coupling $g_0$. To dynamically engineer dissipation and accelerate the qubit decay rate on demand, we employ parametric frequency modulation of the transmon \cite{Maurya_2024_OnDemandDissipation}. By applying an AC flux bias to the transmon's Superconducting Quantum Interference Device (SQUID) loop, the qubit transition frequency becomes time-dependent. To minimize first-order dephasing from flux noise, we bias the transmon at the flux sweet spot. The modulation signal is generated via a standard RF source and gated by a fast microwave switch, resulting in a time-dependent qubit frequency $ \omega_{\mathrm{q}}(t) \approx \omega_{\mathrm{q}} + A_{\mathrm{m}} \cos(\omega_{\mathrm{m}} t)$,
where $\omega_{\mathrm{q}}$ is the static transition frequency, $A_{\mathrm{m}}$ is the effective modulation amplitude, and $\omega_{\mathrm{m}}$ is the modulation frequency. Moving to the interaction picture and utilizing the Jacobi-Anger expansion, this periodic frequency variation dresses the qubit state, generating a series of phase-modulated sidebands. The spectral weight of each sideband is governed by the Bessel functions of the first kind, $J_n(x)$, where the modulation index is $x = A_{\mathrm{m}}/\omega_{\mathrm{m}}$. 

To induce a resonant exchange of energy between the qubit and the readout resonator, the modulation frequency is tuned to bridge the energy gap. In this protocol, we drive the interaction through the $n=2$ sideband, so the modulation frequency is chosen to satisfy $2\omega_{\mathrm{m}} \simeq \omega_{\mathrm{r}}-\omega_{\mathrm{q}}$. Under this condition, the second modulation sideband of the qubit is resonant with the cavity mode. Moving to a frame rotating at the resonator frequency and applying the rotating wave approximation to eliminate fast-oscillating terms, the system dynamics are governed by an effective Hamiltonian $H_{\text{eff}} = g_{\text{eff}} \left( \sigma^+ a + \sigma^- a^\dagger \right)$. Here, $a$ ($a^\dagger$) is the photon annihilation (creation) operator, and $\sigma^-$ ($\sigma^+$) is the qubit lowering (raising) operator. Crucially, the effective coupling strength $g_{\text{eff}}$ is entirely dynamically generated and scales as $g_{\text{eff}} \approx g_0 J_2\left( \frac{A_{\mathrm{m}}}{\omega_{\mathrm{m}}} \right)$. Without this $n=2$ sideband modulation, the resonant coupling $g_{\text{eff}}$ strictly vanishes, and the qubit remains Purcell-protected.

This engineered interaction hybridizes the qubit with the dissipative resonator mode. When the qubit is in the excited state, the $g_{\text{eff}}$ coupling exchanges the excitation with the resonator state. The resonator excitation subsequently decays into the coupled transmission line at the cavity energy decay rate $\kappa$, leading to an enhanced decay via Purcell effect. By digitally gating the modulation signal, we exercise rapid, on-demand control over the qubit's decay rate, providing the tunable non-Hermitian environment required to probe complex dissipative dynamics.

\subsection{Details of the experimental setup}
The experimental platform consists of a flux-tunable superconducting transmon qubit dispersively coupled to a readout resonator with a resonance frequency of $\omega_{\mathrm{r}} / 2\pi = 6.8759$~GHz. During the \textsf{off} segment, the qubit is biased to its sweet spot frequency of $\omega_{\mathrm{q},\textsf{off}} / 2\pi = 4.2713$~GHz. To induce controlled dissipation, we employ a parametric frequency modulation protocol as shown in Fig.~\ref{fig:intro}(a). A microwave source applies the flux-modulation tone to the transmon SQUID loop through a fast flux line (FFL). Temporal control is implemented by passing this modulation tone through a high-speed microwave switch driven by a square-wave gate. The gate defines programmable \textsf{on} durations ($\tau_{\textsf{on}}$) and \textsf{off} durations ($\tau_{\textsf{off}}$), thereby producing the pulsed modulation sequence. This setup allows precise temporal gating of the frequency modulation signal, switching the qubit between a Purcell-protected state and a lossy manifold. During the \textsf{on} segment, the qubit frequency is shifted to $\omega_{\mathrm{q},\textsf{on}} / 2\pi = 4.259$~GHz due to modulation-induced AC Stark shift. This results in resonant modulation frequency $\omega_{\mathrm{m},0} = (\omega_{\mathrm{r}} - \omega_{\mathrm{q},\textsf{on}})/2 \approx 1.308$ GHz. We measure modulation-induced qubit-resonator coupling $g_{\text{eff}}/2\pi = 0.14$~MHz and a resonator decay rate of $\kappa/2\pi = 0.375$~MHz.

\subsection{Continuous and pulsed dissipation}
We first compare the dissipative response produced by continuous modulation with that produced by a periodically gated modulation tone. This comparison establishes the central experimental observation: interrupting the Purcell interaction in time does not simply reduce the average decay rate, but instead reorganizes the frequency response into an interference pattern. To probe this effect, we measure the qubit's excited-state probability ($P_e$) after applying either an uninterrupted modulation tone or a repeated sequence of \textsf{on/off} modulation windows while sweeping the modulation frequency. As illustrated in Fig.~\ref{fig:intro}(b, c), the frequency response undergoes a significant transformation when transitioning from continuous to pulsed parametric modulation. The modulation detuning is defined with respect to resonant modulation frequency $\omega_{\mathrm{m},0}/2\pi = (\omega_{\mathrm{r}}/2\pi - \omega_{\mathrm{q},\textsf{on}}/2\pi)/2$ as $\Delta_{\mathrm{m}}/2\pi = (\omega_{\mathrm{m}} -\omega_{\mathrm{m},0})/2\pi$. In the continuous case, the spectrum is characterized by a single, relatively broad resonance dip centered at the $n=2$ sideband transition, representing an enhanced Purcell decay where the qubit excitation is continuously swapped into the lossy resonator. Conversely, the pulsed case ($\tau_{\textsf{on}} = \tau_{\textsf{off}} = 400$ ns) reveals a series of symmetrically distributed interference dips. Near the center of the spectrum, the interference features are separated by $\delta \omega_{\mathrm{m}}/2\pi = 1/(2t_c)$, where $t_c=\tau_{\textsf{on}}+\tau_{\textsf{off}}$ is the duration of one \textsf{on/off} block. At larger detunings, a weaker set of off-center features appears with twice this spacing. Figure~\ref{fig:intro}(c) extends the measurement of $P_e$ as a function of both modulation detuning $\Delta_{\mathrm{m}}/2\pi$ and the symmetric pulse duration $\tau=\tau_{\textsf{on}}=\tau_{\textsf{off}}$. The resulting two-dimensional spectrum shows that both the central interference dips and the weaker detuned features shift systematically with $\tau$.

\subsection{The Temporal Multi-Slit Analogy}
\label{sec:multislit_analogy}
The periodic nature of the pulsed dissipation sequence invites a direct analogy to the $N$-slit Fraunhofer diffraction. For a diffraction grating of $N$ identical slits of width $a$ and center-to-center spacing $d$, the far-field intensity can be written, up to an overall scale, as $I(\theta) \propto \mathrm{sinc}^2(\beta)\left[\frac{\sin(N\alpha)}{\sin\alpha}\right]^2$ with 
$\beta = \frac{\pi a}{\lambda}\sin\theta$ and $\alpha = \frac{\pi d}{\lambda}\sin\theta$.
 The $\mathrm{sinc}^2(\beta)$ term gives the single-slit envelope and the second factor gives the $N$-slit interference comb.

In this configuration, each \textsf{on} pulse of duration $\tau_{\textsf{on}}$ acts as a temporal slit, during which the qubit state is coherently rotated and subjected to Purcell decay. The \textsf{off} duration $\tau_{\textsf{off}}$ corresponds to the opaque region between slits, where the system accumulates a relative phase but undergoes no transition. The resulting qubit-to-resonator transfer probability derived via the Chebyshev formalism is the temporal equivalent of the grating equation. This mapping demonstrates that pulsed dissipation engineering is essentially the construction of a diffraction grating in the time domain, where the slits are engineered non-Hermitian interactions. In the following section, we make this correspondence explicit by constructing the \textsf{on/off} block propagator and evaluating its repeated action over $N$ cycles. This analytical treatment yields both the survival probability $P_e(N)$ and the expected spacing of the interference dips in the modulation-frequency spectrum.

\section{Results}
\subsection{Analytical model for pulsed dissipation}
We consider an effective two-level model within the single-excitation manifold of a qubit-resonator system
\(
\{|e,0\rangle,\;|g,1\rangle\}
\),
where \(|e,0\rangle\) denotes the qubit excited state with no resonator photon, and \(|g,1\rangle\) denotes the qubit ground state with one resonator photon. During the \textsf{on} segment, the qubit is coupled to the resonator and can decay through the resonator at rate \(\kappa\). During the \textsf{off} segment $g_{\textsf{off}} \approx 0$ as the qubit is effectively decoupled from the resonator. The \textsf{off}-segment Hamiltonian is non-Hermitian due to the continued decay of the resonator photon, but the qubit sector is Purcell-protected from this decay. In this sense, the protocol can be viewed as pulsing the Purcell decay \textsf{on} and \textsf{off} in time.
\label{sec:onoff_derivation}
\begin{figure}[th]
  \centering
  \includegraphics[width=0.9\linewidth]{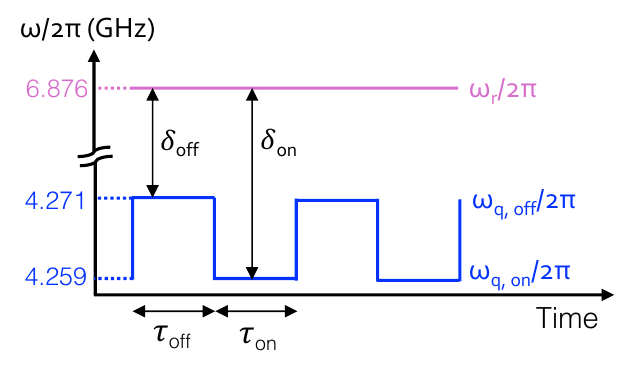}
  \caption{Qubit-resonator dynamics is governed by two different detunings during \textsf{on} and \textsf{off} periods.} 
  \label{fig:delta_onoff}
\end{figure}

\begin{figure*}[t]
  \centering
  \includegraphics[width=.8\textwidth]{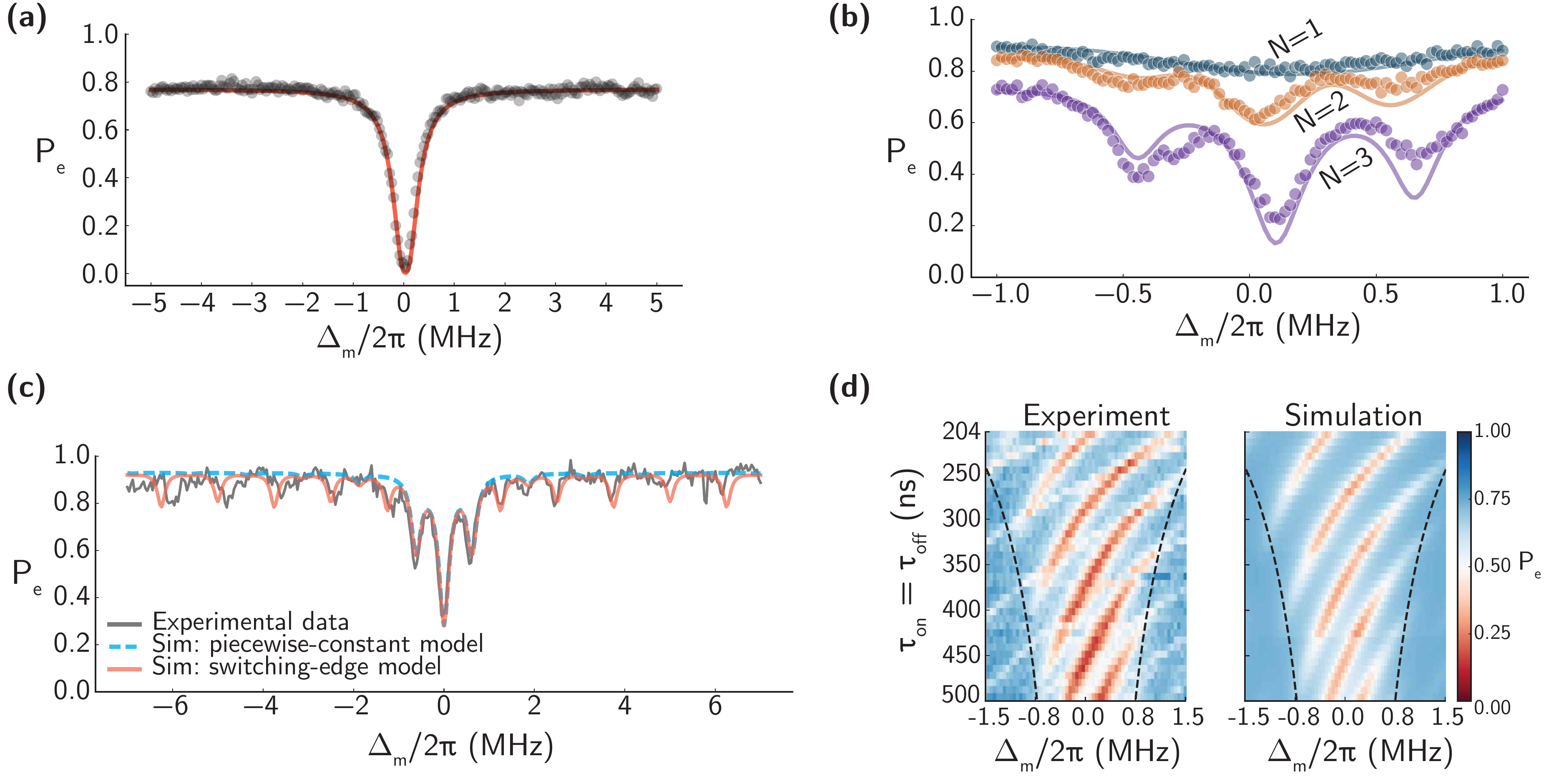}
  \caption{{\bf Comparison between experimental data and numerical simulations based on Eq.~(\ref{eq:Pe_chebyshev}).} Filled circles denote measured excited-state probabilities, while solid lines denote numerical simulations based on the repeated-block expression for $P_e$ in Eq.~(\ref{eq:Pe_chebyshev}). (a) Continuous modulation response, showing a single Purcell-enhanced decay feature near the sideband resonance. (b) Pulsed modulation response for $\tau_{\textsf{on}} = \tau_{\textsf{off}}$ = 440 ns with different numbers of \textsf{on/off} blocks $N$, illustrating the buildup of interference features as the sequence is repeated. (c) Representative pulsed-modulation spectrum showing the two families of spectral features: prominent, central interference dips near $\Delta_{\mathrm{m}}/2\pi=0$ and smaller, off-resonant features at larger detunings. Comparison between experimental data (solid black line) and numerical simulations show that these smaller dips are consequence of sharp \textsf{on/off} switching edges (solid red line) which are not captured with the piecewise-constant model (cyan dashed) where we assume smooth switching. (d) $P_e$ versus $\tau$ and $\Delta_{\mathrm{m}}$ from experiment (left) and  simulation (right), showing interference features confined within the single-slit, $\mathrm{sinc}^2 \left(\frac{\Omega\tau_{\textsf{on}}}{2}\right)$ envelope (dashed black curves) derived in Appendix \ref{app:multislit_diffraction}.}

  \label{fig:expdata}
\end{figure*}

The pulsed modulation sequence is described by alternating \textsf{on/off} Hamiltonians
\begin{equation}
H_{\textsf{on/off}} =
\begin{pmatrix}
0 & g_{\textsf{on/off}} \\
g_{\textsf{on/off}} & \Delta_{\textsf{on/off}}
\end{pmatrix},
\label{eq:H_onoff}
\end{equation}
with $\Delta_{\textsf{on/off}} = \delta_{\textsf{on/off}} - i\frac{\kappa}{2}$.
Here \(g\) is the parametrically activated qubit--resonator coupling during the \textsf{on} window, $\delta_{\textsf{on/off}}= n\omega_{\mathrm{m}}-\left(
\omega_{\mathrm{r}}-\omega_{\mathrm{q},\textsf{on/off}}\right)$ are corresponding sideband detunings in the \textsf{on/off} segments, and \(\kappa\) is the resonator decay rate. In the \textsf{off} window we take \(g_{\textsf{off}} \approx 0\), so that no additional qubit--resonator transfer is generated, while any resonator population already present continues to acquire phase and decay.

The propagator for one \textsf{on/off} block is 
\begin{equation}
M = U_{\textsf{off}}U_{\textsf{on}} =
\begin{pmatrix}
u_{11} & u_{12}\\
\eta u_{21} & \eta u_{22}
\end{pmatrix},
\end{equation}
where $U_{\textsf{on/off}}=e^{-iH_{\textsf{on/off}}\tau_{\textsf{on/off}}}$ and $\eta=e^{-i\Delta_{\textsf{off}}\tau_{\textsf{off}}}$.
For an initial state \(|\psi_0\rangle=|e,0\rangle\), the excited-state
probability after \(N\) \textsf{on/off} blocks $P_e(N)$ is calculated using the Chebyshev-polynomial representation of powers of a \(2\times2\)
matrix,
\begin{equation}
P_e(N) = \left|\mathcal{D}^{(N-1)/2}U_{N-1}(x)u_{11} - \mathcal{D}^{N/2}U_{N-2}(x)\right|^2.
\label{eq:Pe_chebyshev}
\end{equation}
Here \(U_n(x)\) is the Chebyshev polynomial of the second kind,
$\mathcal{D}=\eta e^{-i\Delta_{\textsf{on}}\tau_{\textsf{on}}}$, 
$x = \frac{u_{11}+\eta u_{22}}{2\sqrt{\mathcal{D}}}$.
The details of this derivation, including the explicit forms of \(u_{11}\) and \(u_{22}\), are given in Appendix~\ref{app:chebyshev_derivation}.

The analytical basis for the temporal multi-slit analogy introduced in Sec.~(\ref{sec:multislit_analogy}) is developed explicitly in Appendix~\ref{app:multislit_diffraction}. There we establish the correspondence between pulsed dissipation and Fraunhofer diffraction by showing that the transfer probability separates into two factors: one describing the response of a single modulation window and one describing interference among the $N$ repeated windows. In this comparison, elements of the \textsf{on/off} block matrix $M$ set the broad spectral envelope and the Chebyshev factors encode the repeated application of the block and play the role of the grating factor. Because the block evolution is non-Hermitian, this grating factor is loss-weighted rather than purely unitary, but it still captures the constructive and destructive interference responsible for the sharp dips observed in the spectra.

Figure~\ref{fig:expdata} compares this model directly with the measured dissipation spectra. The continuous-modulation data in Fig.~\ref{fig:expdata}(a) provide the reference response, showing a single Purcell-enhanced decay feature near the sideband resonance. Figure~\ref{fig:expdata}(b) shows that when the modulation is pulsed, the same repeated-block model based on Eq.~(\ref{eq:Pe_chebyshev}) reproduces the emergence and growth of multiple interference dips as the number of \textsf{on/off} blocks is increased. The agreement is also visible in the two-dimensional comparison, where the simulated spectra capture the evolution of the central interference features with pulse duration and modulation detuning. In Fig.~\ref{fig:expdata}(d), the black dashed curves mark the single-window envelope derived in Appendix~\ref{app:multislit_diffraction}; the interference fringes appear within this envelope and shift as the pulse duration is varied. This shows that the dominant structure of the pulsed-dissipation spectrum is set by the combination of the single-window response and the repeated non-Hermitian \textsf{on/off} evolution encoded in the block propagator.

\subsection{Spacing between interference dips}
\label{sec:dip_spacing}

The spacing between the central dips can be estimated from the relative
phase between transfer amplitudes generated during successive
\textsf{on} windows. Since a finite $\kappa$ changes the contrast and width of the interference features but does not affect their spacing, it is not included in this argument. Let the difference between real sideband detunings during the two segments as $d= \delta_{\textsf{off}} - \delta_{\textsf{on}} = \omega_{\mathrm{q},\textsf{off}} - \omega_{\mathrm{q},\textsf{on}}$. The phase difference accumulated between transfer amplitudes generated in two successive cycles is 
$ \Phi = \delta_{\textsf{on}}\tau_{\textsf{on}} + \delta_{\textsf{off}}\tau_{\textsf{off}}$.

Substituting $\delta_{\textsf{off}}=\delta_{\textsf{on}}+d$ and
$\delta_{\textsf{on}}=n\Delta_{\mathrm{m}}$ gives
\begin{equation}
\Phi = n\Delta_{\mathrm{m}}t_c + d\tau_{\textsf{off}}.
\label{eq:cycle_phase}
\end{equation}
The second term is constant for fixed modulation amplitude and pulse
timings and does not affect interference spacing. Constructive transfer occurs when the contributions from successive cycles are integer multiples of $2\pi$ and it appears as a dip in $P_e$. Adjacent dips correspond to a change
$\delta\Phi=2\pi$. From Eq.~\eqref{eq:cycle_phase},
\begin{equation}
\delta\Phi = nt_c\,\delta\Delta_{\mathrm{m}} = 2\pi,
\end{equation}
which, for $n=2$, gives the observed interference dip spacing
\begin{equation}
\delta(\Delta_{\mathrm{m}}/2\pi) = \frac{1}{2t_c}.
\end{equation}
Thus, the central-dip spacing is set by the cycle time. The
single-window coupling and resonator loss primarily determine the
envelope, width, and contrast, while the full dependence on $g$ and
$\kappa$ is captured by Eq.~\eqref{eq:Pe_chebyshev}.

\subsection{Switching-edge model}
\label{sec:switching}

The piecewise-constant model described in the previous sections treats the
\textsf{on} and \textsf{off} segments independently while neglecting
additional dynamics generated at their boundaries. Although this model
captures the central interference dips, it does not reproduce the weaker
outer dips observed at larger modulation detuning. We attribute these
features phenomenologically to spectral components generated by the sharp, 
periodic switching of the modulation waveform. The switching sequence repeats at frequency $\omega_c = \frac{2\pi}{t_c}$.
Periodic switching of a carrier at $\omega_{\mathrm{m}}$ produces
additional modulation components at
$\omega_{\mathrm{m}}+k\omega_c$, where $k$ is an integer. Each component
can activate the $n$th-order parametric sideband when $n\left(\omega_{\mathrm{m}}+k\omega_c\right)
\simeq
\omega_{\mathrm{r}}-\omega_{\mathrm{q},\textsf{on}}$. We model the resulting resonances as an additional weak Purcell-like loss
channel. The survival probability is written as
\begin{equation}
P_e(\Delta_{\mathrm{m}})
=
P_{e,\mathrm{smooth}}(\Delta_{\mathrm{m}})
\exp\left[
-\Gamma_{\mathrm{sw}}(\Delta_{\mathrm{m}})T_{\mathrm{seq}}
\right],
\label{eq:Pe_switching}
\end{equation}
where $P_{e,\mathrm{smooth}}$ is the result of the piecewise-constant model (Eq.~\eqref{eq:Pe_chebyshev}) and $T_{\mathrm{seq}}=Nt_c$ is the total duration of the \textsf{on/off}
sequence. The switching-induced decay rate is taken to be
\begin{equation}
\Gamma_{\mathrm{sw}}(\Delta_{\mathrm{m}})
=
\sum_{k}
|c_k|^2
\frac{g_b^2\kappa}
{
n^2\left(\Delta_{\mathrm{m}}+k\omega_c\right)^2
+\left(\kappa/2\right)^2
}.
\label{eq:Gamma_switching}
\end{equation}
Here $g_b$ is the effective coupling strength associated with the
switching-induced modulation components, while $|c_k|^2$ gives their
relative spectral weights. To determine $c_k$, we approximate the boundary response within one cycle by short transients, one each at the \textsf{on} and \textsf{off} edges. In the impulse approximation, the boundary response is $s_{\mathrm{edge}}(t) = A_{\mathrm{on}}\delta(t) - A_{\mathrm{off}}
\delta\left(t-\tau_{\mathrm{on}}\right)$, where $A_{\mathrm{on}}$ and $A_{\mathrm{off}}$ are complex amplitudes describing the \textsf{on/off} transients. The relative minus sign represents the opposite slopes of the rising and falling edges.
The $k$th Fourier coefficient of such a periodically repeated boundary
response is
\begin{equation}
c_k = A_{\mathrm{on}} - A_{\mathrm{off}}
\exp\left(-ik\omega_c\tau_{\mathrm{on}}\right).
\label{eq:ck}
\end{equation}

\begin{figure*}[t]
  \centering
  \includegraphics[width=.9\linewidth]{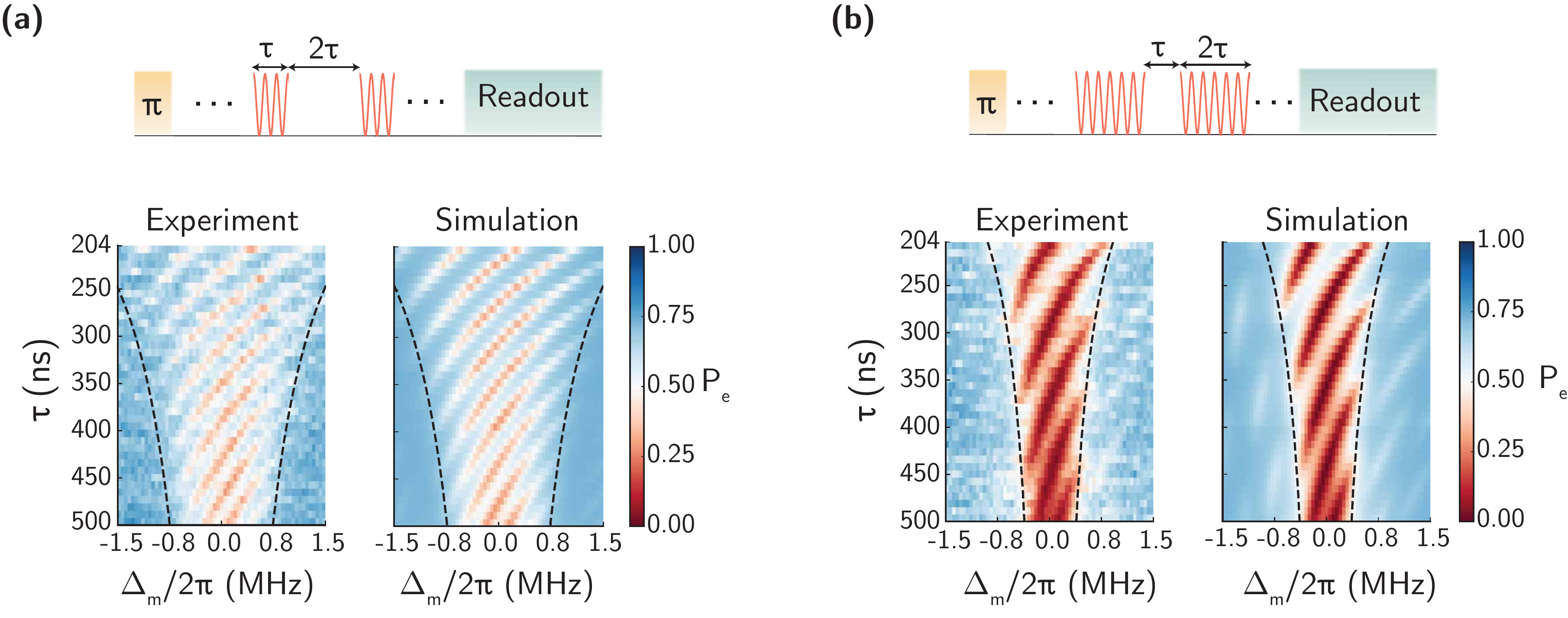}
  \caption{{\bf Effect of duty-cycle variation on dissipation spectra.}
  Measured and simulated two-dimensional maps of the $P_e$ as a function of modulation detuning and pulse duration for two asymmetric \textsf{on/off} sequences. Experimental data and simulation of (a) longer-\textsf{off} sequences, $(\tau_{\textsf{on}},\tau_{\textsf{off}})=(2\tau,\tau)$, and (b) longer-\textsf{on} sequences, $(\tau_{\textsf{on}},\tau_{\textsf{off}})=(\tau,2\tau)$. Because both cases have the same total cycle time $t_c=\tau_{\textsf{on}}+\tau_{\textsf{off}}$, the interference spacing is similar, while the fringe contrast and single-slit envelope (dashed black curves) is different in both cases.}
  \label{fig:duty_cycle}
\end{figure*}

Figure~\ref{fig:expdata}(c) compares the experimental data with the
simulated $P_e(\Delta_{\mathrm{m}}/2\pi)$ obtained from the
piecewise-constant model and the
switching-edge model (Eq.~\eqref{eq:Pe_switching}). The
piecewise-constant model reproduces the central interference dips but
does not capture the weaker dips at larger detuning. Including the
switching-induced loss channels reproduces the positions of both the
central and outer dips. In the numerical simulations based on switching-edge model, $A_{\mathrm{on}}$ is used as the reference amplitude, while
$A_{\mathrm{off}}$ is parameterized by a relative amplitude and phase. Values of $g_b$ and $A_{\mathrm{off}}$ are adjusted to reproduce the overall depth of the outer dips. The sum over $k$ is truncated at a finite value. The switching-edge model is therefore not intended as a microscopic description of the pulse electronics, but as a phenomenological description of the observed weaker, off-resonant resonances. Moreover, these resonances can be suppressed by using smoother pulse edges, which reduce the higher-frequency Fourier components generated by abrupt switching.

\subsection{Effect of duty cycle on interference}
The duty cycle specifies the fraction of each \textsf{on/off} block during which the Purcell interaction is active, $\tau_{\textsf{on}}/(\tau_{\textsf{on}}+\tau_{\textsf{off}})$. Here, we change this fraction while keeping the total block duration fixed at $t_c = \tau_{\textsf{on}} + \tau_{\textsf{off}}$, allowing the effect of the \textsf{on}-window duration to be separated from the overall interference period. Figure~\ref{fig:duty_cycle} shows dissipation spectra obtained with two different duty cycles $(\tau_{\textsf{on}},\tau_{\textsf{off}})=(2\tau,\tau)$ and $(\tau,2\tau)$. Because the total cycle time $(t_c=3\tau)$ is the same in both cases, the interference dips have the same frequency spacing; the main difference appears in their width and contrast. This difference follows from the single-window response: increasing $\tau_{\textsf{on}}$ lengthens the temporal aperture and therefore narrows the frequency response associated with an individual dissipation window. As a result, the $(2\tau,\tau)$ sequence produces a narrower envelope, so fewer interference dips fit within the central response, but the longer interaction time makes the visible dips stronger. Conversely, the \((\tau,2\tau)\) sequence produces a broader envelope, allowing more dips to appear within the same central spacing, but each dip is weaker because the qubit-resonator interaction time per cycle is shorter. Thus the duty cycle provides a direct control knob over the balance between coherent exchange and phase accumulation, thereby shaping both the contrast and spacing of the interference pattern.

\section{Conclusion}

We investigated pulsed Purcell dissipation of a superconducting qubit and the observed spectra are naturally interpreted as a temporal analogue of multi-slit diffraction. Each \textsf{on} window acts as a temporal aperture during which the qubit can exchange excitation with the lossy resonator, while each \textsf{off} window allows relative phase to accumulate without resonant transfer. The resulting interference organizes the dissipative response into a structured spectrum whose characteristic spacing is controlled primarily by the total \textsf{on/off} block duration. A Chebyshev-propagator description captures this repeated-block evolution and connects the repeated pulse train to a grating-like interference factor.

These results show that the temporal structure of engineered dissipation provides a useful control knob for shaping open-system dynamics. By varying the block period and duty cycle, one can tune the spacing, contrast, and envelope of dissipative spectral features without changing the underlying device parameters. This establishes pulsed modulation as a practical extension of dissipation engineering, where the timing of the loss channel becomes an additional experimental parameter.

The results presented here rely on a single Floquet-modulated qubit-resonator pair, in which temporal interference alone shapes the dissipative dynamics. A natural and considerably richer extension is to many-body arrays, where qubits coupled to a common structured reservoir would exhibit both spatial interference, familiar from sub and superradiant emission \cite{van2013photon, kannan2020waveguide, du2026programmable}, and the temporal interference demonstrated in this work. Independently programming the \textsf{on/off} modulation at each site would then realize a genuine spatiotemporal dissipation grating, opening a route to dynamically reconfigurable many-body dissipative phases, directional photon routing, and non-Hermitian Floquet physics \cite{gunderson2021floquet} that are inaccessible with static reservoir engineering

\textit{Acknowledgments}--- We acknowledge Katha Haldar and Yogesh Joglekar, \c{S}ahin \"Ozdemir, Ramy El-Ganainy, and Qi Zhong for discussions. This research was supported by NSF Grant No.~PHY-2408932
, 
the Air Force Office of Scientific Research (AFOSR) Multidisciplinary University Research Initiative (MURI) Award on Programmable systems with non-Hermitian quantum dynamics (Grant No. FA9550-21-1-0202), 
and ONR Grant No.~N000142512160. 
Devices were fabricated and provided by the Superconducting Qubits at Lincoln Laboratory (SQUILL) Foundry at MIT Lincoln Laboratory, with funding from the Laboratory for Physical Sciences (LPS) Qubit Collaboratory.

\bibliography{ref}

\begin{appendix}

\section{Derivation of the excited-state probability $(P_e)$}
\label{app:chebyshev_derivation}

This appendix provides the analytical model used to interpret the pulsed Purcell spectra discussed in the main text. We derive an analytical expression for qubit's excited-state probability after the evolution generated by $N$ repeated \textsf{on/off} modulation blocks. This derivation makes explicit how coherent phase accumulation, qubit--resonator transfer, and resonator decay combine to produce the observed interference structure.

\subsection*{\textsf{on}-segment propagator}
We first consider the Hamiltonian governing the \textsf{on} segment, described in Eq.~(\ref{eq:H_onoff}) of the main text,
\begin{equation}
H_{\textsf{on}} = \begin{pmatrix}
0 & g_{\textsf{on}}\\
g_{\textsf{on}} & \Delta_{\textsf{on}}
\end{pmatrix},
\end{equation}
with $\Delta_{\textsf{on/off}} = \delta_{\textsf{on/off}} - i\frac{\kappa}{2}$ and $\delta_{\textsf{on/off}}= n\omega_{\mathrm{m}}-\left(
\omega_{\mathrm{r}}-\omega_{\mathrm{q},\textsf{on/off}}\right)$.
The propagator for this segment is $U_{\textsf{on}} = e^{-iH_{\textsf{on}}\tau_{\textsf{on}}}$.
We split \(H_{\textsf{on}}\) into trace and traceless parts:
$H_{\textsf{on}} = \frac{\Delta_{\textsf{on}}}{2}I + H'$,
with
\begin{equation}
H'=
\begin{pmatrix}
-\Delta_{\textsf{on}}/2 & g_{\textsf{on}}\\
g_{\textsf{on}} & \Delta_{\textsf{on}}/2
\end{pmatrix}.
\end{equation}
Since \(H'\) is traceless, we have
$(H')^2=\left(\frac{\Delta_{\textsf{on}}^2}{4}+g_{\textsf{on}}^2\right)I$.
It is convenient to define
$\Omega \equiv \sqrt{\Delta_{\textsf{on}}^2+4g_{\textsf{on}}^2}$, so that
$(H')^2=\frac{\Omega^2}{4}I$.
Using the standard exponentiation formula for a traceless \(2\times2\) matrix,
we obtain
\begin{align}
&U_{\textsf{on}} = e^{-i\Delta_{\textsf{on}}\tau_{\textsf{on}}/2} \\
&\left[\cos\!\left(\frac{\Omega\tau_{\textsf{on}}}{2}\right)I
-\frac{2i}{\Omega}\sin\!\left(\frac{\Omega\tau_{\textsf{on}}}{2}\right)
\begin{pmatrix}
-\Delta_{\textsf{on}}/2 & g_{\textsf{on}}\\
g_{\textsf{on}} & \Delta_{\textsf{on}}/2
\end{pmatrix}
\right].
\end{align}

Writing this explicitly,
\begin{equation}
U_{\textsf{on}}=
\begin{pmatrix}
u_{11} & u_{12}\\
u_{21} & u_{22}
\end{pmatrix},
\label{eq:u_explicit}
\end{equation}
with matrix elements
\begin{align*}
&u_{11} = e^{-i\Delta_{\textsf{on}}\tau_{\textsf{on}}/2}
\left[
\cos\!\left(\frac{\Omega\tau_{\textsf{on}}}{2}\right)
+i\frac{\Delta_{\textsf{on}}}{\Omega}
\sin\!\left(\frac{\Omega\tau_{\textsf{on}}}{2}\right)
\right],
\\[4pt]
&u_{22} = e^{-i\Delta_{\textsf{on}}\tau_{\textsf{on}}/2}
\left[
\cos\!\left(\frac{\Omega\tau_{\textsf{on}}}{2}\right)
-i\frac{\Delta_{\textsf{on}}}{\Omega}
\sin\!\left(\frac{\Omega\tau_{\textsf{on}}}{2}\right)
\right],
\\[4pt]
& u_{12}=u_{21} =
-i\,e^{-i\Delta_{\textsf{on}}\tau_{\textsf{on}}/2}
\frac{2g_{\textsf{on}}}{\Omega}
\sin\!\left(\frac{\Omega\tau_{\textsf{on}}}{2}\right).
\end{align*}
\subsection*{\textsf{off}-segment propagator}
During the \textsf{off} segment of duration \(\tau_{\textsf{off}}\), we assume $g_{\textsf{off}}= 0$ and the qubit is decoupled from the resonator, so no additional qubit-to-resonator transfer occurs. However, any population already in the resonator continues to decay at rate $\kappa$. The \textsf{off}-segment propagator is:
\begin{equation}
U_{\textsf{off}}
=
e^{-iH_{\textsf{off}}\tau_{\textsf{off}}}
=
\begin{pmatrix}
1 & 0\\
0 & \eta
\end{pmatrix},
\qquad
\eta \equiv e^{-i \tau_{\textsf{off}} \Delta_{\textsf{off}}}.
\end{equation}
Thus the \textsf{off} window contributes only a relative phase accumulation between the two basis states.

\subsection*{Evolution over repeated \(N\) blocks}
A full \textsf{on/off} block $M = U_{\textsf{off}}U_{\textsf{on}}$ is
\begin{equation}
M=
\begin{pmatrix}
u_{11} & u_{12}\\
\eta u_{21} & \eta u_{22}
\end{pmatrix}.
\end{equation} 
Next we combine the single-block propagators to obtain the excited-state probability after $N$ repeated \textsf{on/off} modulation blocks.
Let the initial state be
$|\psi_0\rangle=
\begin{pmatrix}
1\\
0
\end{pmatrix}$,
corresponding to the qubit initially in \(|e,0\rangle\). After \(N\) blocks, $|\psi_N\rangle = M^N |\psi_0\rangle$. For any \(2\times2\) matrix \(M\), its $N^{th}$ power can be calculated as
\begin{equation}
M^N = \mathcal{D}^{(N-1)/2}U_{N-1}(x)\,M -
\mathcal{D}^{N/2}U_{N-2}(x)\,I,
\label{eq:MtoN}
\end{equation}
where
$\mathcal{D}=\det(M)$, $x=\frac{\mathrm{Tr}(M)}{2\sqrt{\mathcal{D}}}$, and \(U_n(x)\) is the Chebyshev polynomial of the second kind.
For the block matrix $M$, these quantities are
\begin{subequations}
\label{eq:Dandx}
\begin{align}
& \mathcal{D}
= \det(U_{\textsf{off}}) \cdot \det(U_{\textsf{on}})
= \eta\,e^{-i\Delta_{\textsf{on}}\tau_{\textsf{on}}},\\
& x =\frac{u_{11}+\eta u_{22}}{2\sqrt{\mathcal{D}}}.
\end{align}
\end{subequations}
Using Eqs.~\eqref{eq:MtoN} and \eqref{eq:Dandx} to calculate $M^N$ and then taking its \((1,1)\) element, we obtain the excited-state amplitude
$A_e(N) = \langle e,0|M^N|e,0\rangle = (M^N)_{11}$,
and the corresponding probability as
\begin{equation}
P_e(N)= \left|
\mathcal{D}^{(N-1)/2}U_{N-1}(x)\,u_{11}
-
\mathcal{D}^{N/2}U_{N-2}(x)\right|^2
\end{equation}
Thus the final \(P_e(N)\) results from the interplay of non-Hermitian loss during the \textsf{on} segments and coherent phase accumulation during the \textsf{off} segments.
\section{Correspondence to spatial multi-slit diffraction}
\label{app:multislit_diffraction}

Here we connect the temporal \textsf{on/off} modulation sequence to the structure of a spatial diffraction grating. The calculation shows how the transfer amplitude after $N$ identical blocks separates into a single-window envelope and a finite-$N$ interference factor.
The qubit-to-resonator population transfer at the end of the sequence is obtained from the $(2,1)$ element of $M^N$, using Eqs.~\eqref{eq:MtoN} and \eqref{eq:Dandx}:
\begin{equation}
P_{\mathrm{r}}(N)= |\langle g,1|M^N|e,0\rangle|^2 =
|\eta u_{21}|^2 |\mathcal D|^{N-1} |U_{N-1}(x)|^2.
\label{eq:p_r}
\end{equation}
The first factor in Eq.~(\ref{eq:p_r}) is determined by the transfer amplitude during one \textsf{on} window, given by the explicit matrix element in Eq.~(\ref{eq:u_explicit}):
\begin{align}
  |\eta u_{21}|^2 &=|\eta|^2 e^{-\kappa \tau_{\textsf{on}}/2} \left|\frac{2g_{\textsf{on}}}{\Omega}
\sin\left(\frac{\Omega\tau_{\textsf{on}}}{2}\right) \right|^2 \\
&= |\eta|^2 e^{-\kappa \tau_{\textsf{on}}/2} g_{\textsf{on}}^2 \tau_{\textsf{on}}^2
\left| \mathrm{sinc}
\left(\frac{\Omega\tau_{\textsf{on}}}{2} \right)  \right|^2
\end{align}
To simplify $|U_{N-1}(x)|^2$ factor in Eq.~(\ref{eq:p_r}), we define a normalized matrix $\widetilde{M} \equiv M / \sqrt{\mathcal{D}}$, ensuring $\det(\widetilde{M}) = 1$. The trace of this normalized matrix relates to the Chebyshev parameter $x$ via
\begin{equation}
\text{Tr}(\widetilde{M}) = \frac{\text{Tr}(M)}{\sqrt{\mathcal{D}}} = 2x.
\end{equation}
The characteristic equation of $\widetilde{M}$ matrix, $\lambda^2 - 2x\lambda + 1 = 0$, yields the system eigenvalues $\lambda_{\pm} = x \pm \sqrt{x^2 - 1}$. Although $x$ is generally complex due to resonator decay $\kappa$, the unimodular constraint $\lambda_+ \lambda_- = 1$ permits the parameterization $\lambda_{\pm} = e^{\pm i\alpha}$ via a complex phase variable $\alpha \equiv \alpha_{\mathrm{r}} + i\alpha_{\mathrm{i}}$. Evaluating the trace as $\text{Tr}(\widetilde{M}) = \lambda_+ + \lambda_-$ yields $x = \cos\alpha$.
Using the standard relation, $U_{N-1}(\cos\alpha) = \frac{\sin(N\alpha)}{\sin\alpha}$, satisfied by the Chebyshev polynomial of the second kind, in Eq.~(\ref{eq:p_r}),
the transfer probability becomes
\begin{equation}
P_{\mathrm{r}}(N) \propto |\mathcal D|^{N-1} \mathrm{sinc}^2 \left(\frac{\Omega\tau_{\textsf{on}}}{2}\right) \left|\frac{\sin(N\alpha)}{\sin\alpha} \right|^2.
\end{equation}
This expression has the similar structure as the Fraunhofer diffraction
formula for an \(N\)-slit grating:
\begin{equation}
I_{\mathrm{opt}}(\theta_{}) \propto \mathrm{sinc}^2(\beta)
\left[
\frac{\sin(N\alpha)}
{\sin\alpha}
\right]^2.
\end{equation}
Thus, the $\mathrm{sinc}^2\left(\frac{\Omega\tau_{\textsf{on}}}{2}\right)$ term plays the role of the single-slit envelope, while the ratio $\frac{\sin(N\alpha)}{\sin\alpha}$ is the finite-\(N\) interference factor. The additional prefactor \(|\mathcal D|^{N-1}\) accounts for attenuation over repeated non-Hermitian blocks. 
\end{appendix}

\end{document}